\documentclass[twocolumn,showpacs,preprintnumbers,amsmath,amssymb,nofootinbib]{revtex4}

\usepackage{graphicx}
\usepackage{dcolumn}
\usepackage{bm}

\newcommand{\be}{\begin{equation}}
\newcommand{\ee}{\end{equation}}
\newcommand{\ba}{\begin{eqnarray}}
\newcommand{\ea}{\end{eqnarray}}

\begin{document}

\preprint{CERN-PH-TH/2007-077, ROME-1451/2007}

\title{Theta dependence of the vacuum energy in the SU(3) gauge theory from the lattice}

\author{Leonardo Giusti$^{a,}$\footnote{On leave from Centre de Physique Th\'eorique, 
               Case 907, CNRS Luminy, F-13288 Marseille Cedex 9, France.}, 
        Silvano Petrarca$^{b,c}$, Bruno Taglienti$^c$}

\affiliation{\vspace{0.3cm}
\vspace{0.1cm} $^a$ CERN, Department of Physics, TH Division, CH-1211 Geneva 23, 
               Switzerland.\\
               $^b$ Dipartimento di Fisica, Universit\`a di Roma ``La Sapienza'', 
               P.le A. Moro 2, I-00185 Rome, Italy.\\
               $^c$ INFN, Sezione di Roma, P.le A. Moro 2, I-00185 Rome, Italy.}


\begin{abstract}
We report on a precise computation of the topological charge 
distribution in the SU(3) Yang--Mills theory. It is carried out on the 
lattice with high statistics Monte Carlo simulations by employing the definition 
of the topological charge suggested by Neuberger's fermions. We observe significant 
deviations from a Gaussian distribution. Our results disfavour the $\theta$ behaviour of the 
vacuum energy predicted by instanton models, while they are compatible with the 
expectation from the large $N_c$ expansion. 
\end{abstract}

\pacs{
11.15.Pg   
11.15.Ha,  
12.38.Gc 
\vspace{-0.25cm}
}
\maketitle

{\em Introduction.---} The $\theta$ dependence of the vacuum 
energy~\cite{Belavin:1975fg,'tHooft:1976up,Jackiw:1976pf,Callan:1976je}, or 
equivalently the functional form of the topological charge distribution, 
is a distinctive feature of the ensemble of gauge configurations 
that dominate the path integral of a Yang--Mills theory.
In the Euclidean space-time the ground-state energy $F(\theta)$ is 
defined as 
\be
e^{- F(\theta)} = \langle e^{i\theta Q}\rangle\; ,
\ee
where, as usual, $\langle\dots\rangle$ indicates the path-integral 
average (our normalization is $F(0)=0$). In the large volume regime 
$F(\theta)$ is proportional to the size $V$ of the system,
a direct consequence of the fact that the topological charge operator $Q$ 
is the four-dimensional integral of a local density. 
The function $F(\theta)$ is related to the probability 
of finding a gauge field configuration with 
topological charge $Q=\nu$ by the Fourier transform
\be\label{eq:pnu}
P_\nu = \int_{-\pi}^\pi \frac{d\theta}{2 \pi} 
e^{-i\theta\nu} e^{- F(\theta)}\; .
\ee
\indent Large $N_c$ arguments~\cite{'tHooft:1973jz},
with $N_c$ being the number of colors, suggest that the fluctuations 
of the topological charge are of quantum non-perturbative 
nature~\cite{Witten:1978bc,Witten:1980sp}. 
The $\theta$ dependence of the vacuum energy is expected 
at leading order in $1/N_c$, and the normalized cumulants 
\be\label{eq:cn}
{C}_{n} = (-1)^{n+1}\frac{1}{V}\frac{d^{2n}}{d\theta^{2n}} 
F(\theta) \Big|_{\theta=0}\qquad n=1,2,\dots\; ,
\ee
which should scale asymptotically as 
$N_c^{2-2n}$~\cite{Witten:1978bc,Witten:1980sp}, have to be 
determined with a non-perturbative computation. 
On the other hand several models, such as the dilute gas or liquid of instantons, assume
that the path integral is dominated by semiclassical 
configurations~\cite{'tHooft:1976fv,Callan:1977gz,Shuryak:1981ff,Diakonov:1983hh}. 
They predict a $\theta$ behaviour of the form 
\be\label{eq:finst}
F^\mathrm{Inst} (\theta) = -V A \{\cos(\theta)-1\}\;,
\ee 
with $A$ being exponentially suppressed at large $N_c$.\\
\indent The $\theta$ dependence of the vacuum energy plays a 
crucial r\^ole also in the solution of the 
so-called $U(1)_A$ problem in QCD. 
The Witten--Veneziamo mechanism relates the cumulants of the 
topological charge distribution in the Yang--Mills theory
with the leading anomalous contribution to the mass and scattering 
amplitudes of the $\eta'$ 
meson in QCD~\cite{Witten:1979vv,Veneziano:1979ec,Seiler:1987ig,Giusti:2001xh,Seiler:2001je}.
The known value of ${C}_1$ in the SU(3) theory supports indeed 
the fact that the bulk of the $\eta'$ mass is due to 
the anomaly~\cite{DelDebbio:2004ns}.\\
\indent Recent theoretical developments in lattice gauge theory
made it possible to find an unambiguous definition of the topological charge 
distribution with a finite and universal continuum 
limit~\cite{Neuberger:1997fp,Luscher:1998pq,Hasenfratz:1998ri,Giusti:2001xh,Giusti:2004qd,Luscher:2004fu}. 
The aim of this work is a precise computation of the distribution 
of the topological charge in the SU(3) Yang--Mills theory. We observe 
significant deviations from a Gaussian behaviour: they
disfavor the $\theta$ dependence given in Eq.~(\ref{eq:finst}), while
they are compatible with expectations from the large $N_c$ expansion.\\
\indent In the past the distribution of the topological charge was
already studied (see Ref.~\cite{DelDebbio:2004ns} and references therein).
These computations, however,  
were not precise enough to observe deviations from the 
leading Gaussian behaviour. In this respect we have exploited the 
efficiency of the algorithm for the determination of the charge 
developed in Ref.~\cite{Giusti:2002sm}. Properties of 
the charge distribution have been investigated also with fermionic
and bosonic methods (see Refs.~\cite{DelDebbio:2002xa,Durr:2006ky} and references therein). 
These results, however, are affected by model-dependent systematic errors that are not 
quantifiable, and their interpretation rests on a weak theoretical ground. 

{\em Topological charge definition.---} The Neuberger--Dirac operator $D$ is defined as 
\begin{eqnarray}
D & = & \frac{1}{\bar a} \Big[1 + \gamma_5 \mathrm{sign}(H)\Big]\; ,\\
H & = & \gamma_5 (a D_\mathrm{w} -1 -s)\, , \qquad \bar a = \frac{a}{1+s}\, ,
\label{eq:overlap}
\end{eqnarray}
where $D_\mathrm{w}$ is the standard Wilson--Dirac operator and
$s$ is an adjustable parameter in the range $|s|<1$ (for notations not 
explained here see Ref.~\cite{Giusti:2002sm}). It satisfies 
the Ginsparg--Wilson relation \cite{Ginsparg:1982bj}, and therefore
the associated fermion action preserves an exact chiral 
symmetry at finite lattice spacing~\cite{Luscher:1998pq}.
The corresponding Jacobian is non-trivial, and the chiral anomaly 
is recovered {\it \`a la} Fujikawa \cite{Luscher:1998pq,Fujikawa:1979ay} with 
the topological charge density operator defined  as~\cite{Hasenfratz:1998ri}
\begin{equation}\label{eq:qx}
a^4 q(x) = -\frac{\bar a}{2}\, \mathrm{Tr}\Big[\gamma_5 D(x,x)\Big] , 
\end{equation}
where the trace runs over spin and color indices. With this 
definition the topological charge in a given background is given by 
$Q \equiv \sum_x q(x) = n_+ - n_-$,  with $n_+$ ($n_-$) being the 
number of zero modes of $D$ with positive (negative) chirality.
The normalized cumulants ${C}_{n}$ are thus defined as the 
integrated connected correlation functions of $n$
charge densities\footnote{Correlation functions
of an odd number of topological charges vanish thanks 
to the invariance of the theory under parity.}
\be\label{eq:CnQCd}
{C}_{n} = \frac{a^{8n}}{V} \sum_{x_1,\dots,x_{2n}}
\langle q(x_1)\dots q(x_{2n})\rangle^\mathrm{con}\; .
\ee
They have an unambiguous finite continuum limit which is 
independent of the details of the 
regularization~\cite{Giusti:2001xh,Giusti:2004qd,Luscher:2004fu}.
At finite lattice spacing they are affected by discretization 
errors which start at $O(a^2)$.

{\em The large volume limit.---} Being $\nu$ integer-valued, 
$F(\theta)$ is a periodic function with period $2\pi$. 
In the interval $-\pi<\theta<\pi$ it has 
its absolute minimum at $\theta=0$, and it may then be 
expanded as 
\be\label{eq:expEt}
F(\theta) = V \sum_{n=1}^{\infty} (-1)^{n+1} 
              \frac{\theta^{2n}}{(2n)!} {C}_{n}\; . 
\ee
Leading finite-size effects in the ${C}_n$ are 
exponentially suppressed at asymptotically large volumes. 
They are proportional to $e^{-M_g L}$,  
with $M_g\sim 1.6$~GeV being the lightest glueball 
mass~\cite{Chen:2005mg}, and they become rapidly negligible 
as soon as $L$ is larger than $1$~fm or so~\cite{Giusti:2003gf}.
By inserting Eq.~(\ref{eq:expEt}) in 
Eq.~(\ref{eq:pnu}), and neglecting exponentially small corrections 
proportional to $e^{-2\pi^2\sigma^2}$, we can express the topological 
charge distribution $P_\nu$ at large volumes by
a saddle point expansion (usually named Edgeworth expansion 
in statistics\footnote{The Edgeworth expansion is usually 
adopted in the context of the central limit theorem~\cite{Cramer:1945}. 
In our case the volume $V$, or $(N_c^2 V)$ at large $N_c$, plays the r\^ole of 
the number of independent degrees of freedom.}). This is 
an asymptotic series in powers of $1/V$ 
(or $1/(N_c^2 V)$ according to large $N_c$)
which, up to higher order corrections, reads 
\be\displaystyle\label{eq:pnuege}
P_\nu = \frac{e^{-\frac{\nu^2}{2 \sigma^2}}}{\sqrt{2 \pi \sigma^2}}  
\left[1 + \frac{1}{4!}\frac{\tau}{\sigma^2} \mathrm{He}_\mathrm{4}\left(\nu/\sigma\right)
\right]\; .
\ee
The parameters are $\sigma^2 = V {C}_1$ and $\tau={C}_2/{C}_1$, and 
the Hermite polynomial $\mathrm{He}_\mathrm{4}$ can be found in Ref.~\cite{Abramowitz:1965}.\\ 
\indent The semiclassical models provide a sharp prediction
of the topological charge distribution. By inserting Eq.~(\ref{eq:finst}) in 
Eq.~(\ref{eq:pnu}), and taking into account that $\nu$ is an integer,
we obtain
\be\label{eq:pnuinst} 
P^\mathrm{Inst}_\nu = e^{-VA} I_\nu(VA)\; ,
\ee
where $I_\nu$ are the modified Bessel functions of the first kind~\cite{Abramowitz:1965}. By
construction all normalized cumulants are equal to $A$.

\begin{figure}[t]
\begin{center}
\includegraphics*[width=8.5cm]{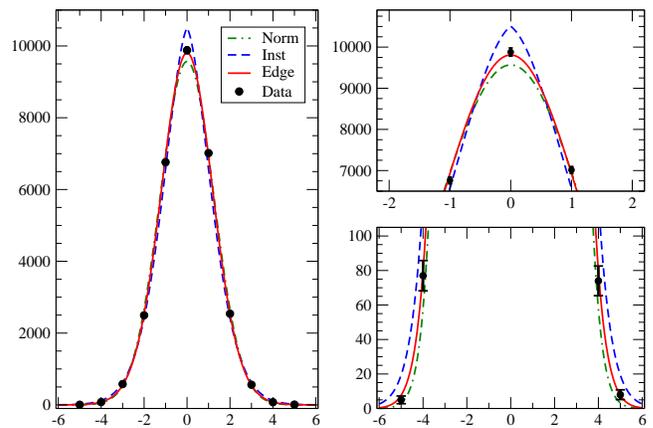}
\vspace{-0.25cm}
\caption{Number of configurations vs the topological charge for the lattice ${\rm A}_2$. To
guide the eye, lines connect the values of the fitted distributions: the simple 
Gaussian (Norm), the instanton prediction in Eq.~(\ref{eq:pnuinst}) (Inst) and the 
Edgeworth expansion in Eq.~(\ref{eq:pnuege}) (Edge). The plots on right are a blowup 
of the top and the bottom of the distribution.\label{fig:A2}}
\end{center}
\end{figure}

\begin{figure}[!t]
\begin{center}
\includegraphics*[width=6.0cm]{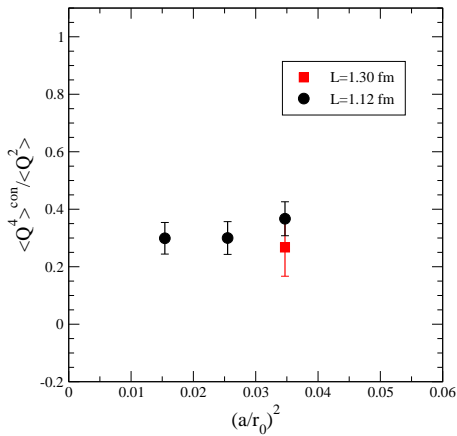}
\vspace{-0.25cm}
\caption{Ratio of the first two cumulants vs the lattice spacing.\label{fig:cont}}
\end{center}
\end{figure}

{\em Lattice computation.---} The numerical computation 
is performed by standard Monte Carlo
techniques. 
The ensembles of gauge configurations are generated 
with the Wilson action and periodic boundary conditions.
Each update cycle consists in 1 heat-bath and several over-relaxations 
of all link variables (more details can be found in Ref.~\cite{Giusti:2003gf}).
The charge density is defined as in 
Eq.~(\ref{eq:qx}) with $s=0.4$, and the corresponding topological
charge has been computed by counting the number of zero modes 
of the Neuberger--Dirac operator with the algorithm proposed 
in Ref.~\cite{Giusti:2002sm}.\\ 
\indent The list of lattices, the 
value of the bare coupling constant $\beta=6/g_0^2$, the linear size $L/a$ 
in each direction,  and the number of independent configurations are reported in 
Table~\ref{tab:lattices}. Lattice spacings and volumes have been chosen to have
normalized cumulants with small discretization and finite-size errors. To 
estimate discretization effects we have simulated three lattices, 
${\rm A}_1$--${\rm A}_3$, with the same physical volume but different lattice spacings.
Two additional lattices, ${\rm B}_1$ and ${\rm C}_1$, have been generated to quantify the 
magnitude of finite-size effects: they have the same bare coupling of ${\rm A}_1$ 
but larger volumes. The autocorrelation 
function of the topological charge has been computed at the three values of $\beta$ 
by monitoring its value for several thousands of consecutive update
cycles of the $A$ lattices. The corresponding autocorrelation time is 
between  $20$ and $40$ cycles. Since the cost of a Monte Carlo update 
is negligible with respect to the computation of the index of the 
Neuberger-Dirac operator, we separated subsequent measurements on all lattices 
by a number of update cycles between 1 and 2 orders of magnitude larger than the above estimates. 
Statistical errors are then computed with the jackknife method by considering the 
measurements as independent. A preliminary analysis of a subset of our results 
was presented at the conference ``Lattice 2006''~\cite{Giusti:2007lat}.\\
\indent The Monte Carlo technique adopted here generates the gauge configurations 
with a probability density proportional to $e^{-S_\mathrm{YM}}$, with 
$S_\mathrm{YM}$ being the chosen discretization of the Yang--Mills action. This 
algorithm performs an importance sampling of the topological charge with the 
probability distribution given in Eq.~(\ref{eq:pnu}). A statistical signal for the 
$n^\mathrm{th}$ cumulant is then obtained only if the number of configurations is 
high enough for the sample to be sensitive to terms suppressed 
as $V^{n-1}$ in the Edgeworth expansion. For instance, the estimators of the first two 
cumulants
\ba
\overline{Q^2} & = & \frac{1}{N}
\sum_{i=1}^{N} \nu^2_i\; , \label{eq:est1}\\[0.125cm]
\overline{Q^{4,}} \,\!^\mathrm{con} & = & 
\frac{1}{N}\sum_{i=1}^{N} \nu^4_i - 3 \left(\frac{1}{N}\sum_{i=1}^{N} \nu^2_i\right)^2\; ,
\label{eq:est2}
\ea
with $\nu_i$ being the value of the topological charge for a given gauge configuration
and $N$ the total number of configurations, have variances which, up to sub-leading 
corrections, are $(2\sigma^4+\sigma^2\tau)/N$ and $(24\sigma^8 + 72 \sigma^6\tau)/N$
respectively.
The number of configurations on our main set of lattices, the ${\rm A}$ series,
has been fixed to have a precision of $15-20\%$ on the second cumulant $C_2$. For the 
lattice ${\rm B}_1$  the number of configurations is chosen so to have a signal for
$\langle Q^4\rangle^\mathrm{con}$ and therefore a rough estimate 
of finite-size effects\footnote{Given the scaling of the statistical error
with $V$ and $n$, it is very inefficient to compute higher cumulants at 
large volumes with the standard sampling procedure. Once $\sigma^2$ is known, however,
one could devise an adaptive importance sampling algorithm which integrates
this information so to have a reduced $\sigma^2_\mathrm{eff}$.}.
Lattice ${\rm C}_1$ has been simulated to quantify finite-size 
effects in $\langle Q^2\rangle$ with confidence.\\
\begin{table}[!t]
\begin{center}
\setlength{\tabcolsep}{.15pc}
\begin{tabular}{llcccclll}
\hline
Lat    &$\beta$&$L/a$&$r_0/a$&$L$[fm]&$N$& $\chi^{2,\mathrm{Norm}}_\mathrm{dof}$ 
&$\chi^{2,\mathrm{Inst}}_\mathrm{dof}$&$\chi^{2,\mathrm{Edge}}_\mathrm{dof}$\\[0.125cm]
\hline
${\rm A}_1$&$6.0$   &$12$& $5.368$ &$1.12$&$34800$& $\;15$  &$\;27$    &$\;1.5$\\[0.125cm]
${\rm A}_2$&$6.0938$&$14$& $6.263$ &$1.12$&$30000$& $\;12$  &$\;34$    &$\;1.4$\\[0.125cm]
${\rm A}_3$&$6.2623$&$18$& $8.052$ &$1.12$&$30000$& $\;13$  &$\;41$    &$\;1.1$\\[0.125cm]
${\rm B}_1$&$6.0$   &$14$& $5.368$ &$1.30$&$30000$& $\;1.3$ &$\;6.7$   &$\;0.14$\\[0.125cm]
${\rm C}_1$&$6.0$   &$16$& $5.368$ &$1.49$&$10000$& $\;0.67$&$\;2.4$   &$\;0.79$\\
\hline
\end{tabular}
\caption{Simulation parameters, number $N$ of configurations generated, 
and values of $\chi^2_\mathrm{dof}$ for the fit of the data to a Gaussian
(Norm), to the instanton prediction in Eq.~(\ref{eq:pnuinst}) (Inst) and to 
the Edgeworth expansion in Eq.~(\ref{eq:pnuege}) (Edge).\label{tab:lattices}}
\end{center}
\end{table}
\indent For each lattice we have compared the histogram of the topological 
        charge distribution with three functional forms: a simple Gaussian 
        centered at the origin, the Edgeworth expansion in Eq.~(\ref{eq:pnuege}), 
        and the prediction from instanton models in Eq.~(\ref{eq:pnuinst}). 
        The free parameter(s) of each function has(ve) been fixed by maximizing 
        the likelihood. For the symmetrized histograms the values of 
        $\chi^2$ per degree of freedom at the minimum are reported in 
        Table~\ref{tab:lattices}, and for the lattice
        ${\rm A}_2$ the data points and the three curves are shown in Fig.~\ref{fig:A2}.  
        The Edgeworth expansion reproduces well the behaviour 
        of the numerical data at these volumes and lattice spacings within our 
        statistical errors\footnote{We also fitted the data with the 
        functional form in Eq.~(\ref{eq:pnu}) and two non-vanishing cumulants. 
        The conclusions are analogous.}. On the ${\rm A}$ lattices the Gaussian distribution is 
        incompatible with the data, while at the two larger volumes the 
        $\chi^2_\mathrm{dof}$ is still rather good. The functional 
        form suggested by instanton models is excluded on the lattices 
        ${\rm A}_1$--${\rm A}_3$ and ${\rm B}_1$, and is off by more than two sigmas 
        on the lattice ${\rm C}_1$. On the ${\rm A}$ lattices a fit limited to $|\nu|\leq 1$ leads 
        to the same conclusions. This is one of the main results of this paper.\\
\indent To quantify the magnitude of discretization and finite-size effects we have 
        computed the first two cumulants of the topological charge distribution  
        with the estimators given in Eqs.~(\ref{eq:est1}) and (\ref{eq:est2}).
        The numerical results are reported in Table~\ref{tab:cumulants}.
        The contributions from the poorly sampled tail of the distributions, 
        i.e. bins of the symmetrized histogram populated by less than 10 events,   
        have been estimated from the large volume expression in Eq.~(\ref{eq:pnuege}) 
        as suggested in Ref.~\cite{Giusti:2003gf}. Within our statistical errors we do 
        not observe a signal for the higher cumulants, and correlation functions 
        of an odd number of topological charges are always compatible with zero.\\ 
\indent For the data samples that can be directly compared, the values of $\langle Q^2\rangle$ 
        are in very good agreement with the results in Ref.~\cite{DelDebbio:2004ns}. 
        We confirm that discretization effects on $r_0^4 {C}_1$, where $r_0$ is a low-energy 
        reference scale well measured in the pure gauge theory~\cite{Guagnelli:1998ud}, 
        are moderate (of the order of $10\%$ at our coarser lattice spacing), and finite size
        effects are below $5\%$ at our smaller volume. Even though our errors 
        for $\langle Q^2\rangle$ are much smaller with respect to those in Ref.~\cite{DelDebbio:2004ns}, 
        a significative improvement of the determination of ${C}_1$ in the continuum limit requires 
        further simulations and is left to a future publication.\\
\indent The second cumulant is best expressed by the adimensional ratio 
        $\langle Q^4\rangle^\mathrm{con}/\langle Q^2\rangle$ which 
        has a well defined continuum and infinite-volume limit. The numerical
        results for the lattices ${\rm A}_1$--${\rm A}_3$ and ${\rm B}_1$ are 
        reported in Table~\ref{tab:cumulants}, and they are plotted in 
        Fig.~\ref{fig:cont} as a function of $(a/r_0)^2$. All values
        are incompatible with $1$, the predicted value from Eq.~(\ref{eq:finst}). 
        The results from ${\rm A}_1$, ${\rm A}_2$ and ${\rm A}_3$ agree 
        within errors. No statistical-significant evidence of 
        discretization effects is thus observed. The theoretical arguments given in 
        the third section suggest small finite-size effects at these volumes.
        A direct estimate of these effects on $\langle Q^4\rangle^\mathrm{con}$ 
        would require more precise data for the lattice ${\rm C}_1$. The compatibility of the 
        results on lattices ${\rm A}_1$ and ${\rm B}_1$, however, is
        consistent with the theoretical expectations. Our best estimate for 
        the ratio of the first two cumulants  
        is $\langle Q^4\rangle^\mathrm{con}/\langle Q^2\rangle=0.30(11)$.
        The central value is taken from the lattice ${\rm A}_3$, the one
        with the finer lattice spacing, and the error is the 
        sum in quadrature of the statistical error and of the difference 
        between the central values computed on the lattice ${\rm A}_1$ 
        and ${\rm B}_1$.
\begin{table}[!t]
\begin{center}
\setlength{\tabcolsep}{.375pc}
\begin{tabular}{llll}
\hline
Lat    &$\;\;\;\langle Q^2\rangle$& $\;\;\;\;\langle Q^4\rangle^\mathrm{con}$ 
       & $\langle Q^4\rangle^\mathrm{con}/\langle Q^2\rangle$\\[0.125cm]
\hline
${\rm A}_1$&$1.637(13)$&$\;\;\;0.60(9)$& $\;\;\;0.37(6)$\\[0.125cm]
${\rm A}_2$&$1.566(13)$&$\;\;\;0.47(9)$& $\;\;\;0.30(6)$\\[0.125cm]
${\rm A}_3$&$1.432(12)$&$\;\;\;0.43(7)$& $\;\;\;0.30(5)$\\[0.125cm]
${\rm B}_1$&$3.09(3)$  &$\;\;\;0.8(3)$ & $\;\;\;0.27(10)$\\[0.125cm]
${\rm C}_1$&$5.44(8)$  &$\;\;\;\;\;\; - $     &$\;\;\;\;\;\; - $\\
\hline
\end{tabular}
\caption{Results for the first two cumulants and their ratio.\label{tab:cumulants}}
\end{center}
\end{table}

{\em Final remarks.---} From the previous analysis we conclude that our 
results disfavour the $\theta$ behaviour of the 
vacuum energy predicted by instanton models.
Our best estimate of the ratio 
$\langle Q^4\rangle^\mathrm{con}/\langle Q^2\rangle = 0.30(11)$
is incompatible with 1, which is the value predicted from Eq.~(\ref{eq:finst}).
This suggests that the quantum fluctuations of the topological charge
are of quantum non-perturbative nature in the ensemble of gauge 
configurations that dominate the path integral.
The large $N_c$ expansion does not provide a sharp prediction for the value 
of $\langle Q^4\rangle^\mathrm{con}/\langle Q^2\rangle$.  Its small value, 
however, is compatible with being a quantity suppressed in the large $N_c$ limit. The value 
of $\langle Q^4\rangle^\mathrm{con}$ is related via the Witten--Veneziano mechanism to the leading anomalous 
contribution to the $\eta'$--$\eta'$ elastic scattering amplitude in QCD.

We thank L. Del Debbio, L. \'Alvarez-Gaum\'e, M.~L\"uscher, M. Testa and G. Veneziano 
for stimulating discussions on the topic of this paper. The numerical computations
have been carried out on PC farms of the 
Italian INFN Grid project. We warmly thank Giuseppe Andronico for his
work in the organization of Theophys, the INFN Grid virtual organization 
for theoretical physics. Many thanks to Alessandro De Salvo and Marco Serra 
for the continuous effort in helping us with the Grid during the 
accomplishment of the project.
\bibliographystyle{apsrev}
\bibliography{Q4}
\end{document}